# Thickness dependent dark exciton emission in $(PEA)_2PbI_4$ nanoflake and its brightening by in-plane magnetic field

Wei Tang[1], Liting Tao[2], Tian Zhang[1], Yanjun Fang[2*], Deren Yang[2], Linjun Li[1*]

1 State Key Laboratory of Modern Optical Instrumentation, College of Optical Science and Engineering,

2 State Key Laboratory of Silicon Materials, School of Materials Science and Engineering,

Zhejiang University, Hangzhou 310027, China.

*Corresponding authors: lilinjun@zju.edu.cn; jkfang@zju.edu.cn

**Halide perovskite materials raised tremendous interest in recent years since their cheap fabrication, superior performance in both solar cell and light emitting diode (LED). Due to the existence of layered quantum well structure, quasi two-dimensional(2D) halide perovskite has more intriguing spin related physics than its 3D counterpart. For instance, the detection and brightening of dark exciton (DX) in 2D halide perovskite attracts much attention since these species can be used in opto-spintronic and quantum computing devices. Here, we report the gradually brightened emission of the DX at 2.33 eV with the thickness decreases in $(PEA)_2PbI_4$ single crystalline nanoflake, which hitherto has not been reported. By coupling with in-plane (IP) magnetic field ($B_\parallel$) in Voigt configuration, the DX emission can be sharply enhanced, while for the out-of-plane (OP) magnetic field ($B_\perp$) in Faraday configuration, the DX emission has no noticeable change, which can be reconciled with the theory interpretation of magnetic field dependent wave function mixing between the four exciton states $\phi_1$, $\phi_2$, $\phi_3^-$, $\phi_3^+$. The emission of DX $\phi_2$ at 2.335 eV and the fine splitting of all the four states are observed in static PL spectroscopy for the first time. Our work thus clarifies the debating questions regarding to previous research on DX behavior in 2D halide perovskite material and sheds light on the road of realizing opto-spintronic or quantum computing devices with these materials.**

**Introduction：**

In solid state physics, lattice structure determined crystal field together with spin orbit coupling, Coulomb interaction, electron hole exchange interaction, etc, renders the crystal owning bundles of exotic electronic properties. Among all the crystalline structures[1], Perovskite type compounds received special attention since they give birth of charge, magnetic order and high temperature superconductivity because that all the elements including charge, spin, orbit, lattice and their interactions can be related to the unique octahedron structure. In the past decade, the halide organic and inorganic Perovskite (HOIP) received reviving attention since they have very high mobility and superior photoelectric properties, which can be used as solar cells and light emitting diodes (LED) material[2-5]. Quasi 2D HOIP has a layered quantum well structure, possessing inherent strong quantum confinement effects and large electron-hole exchange interaction, resulting in a large fine structure splitting of excitons, hence complicates the selection rule governed bright(BX) and dark exciton(DX) behavior in the 2D HOIP[6-8]. DX has a longer spin lifetime and different spin configuration than the bright species and hence can be useful in opto-spintronics and quantum computation devices[9-11]. To improve the efficiency of LED devices, manipulating the DX and its energy difference with the BX is a promising route to transfer DX to BX to promote the total light emission efficiency. Therefore, how to detect or brighten the DX in 2D HOIP attracted much attention in recent years.

In general for 2D HOIP, the crystal field, spin orbit coupling, together with the strong Coulomb exchange interaction, gives the four non degenerate exciton states: ɸ$_1$, ɸ$_2$, ɸ$_3^+$, ɸ$_3^-$, where ɸ$_1$ is the J=0 optically inactive ground state, ɸ$_2$, ɸ$_3^+$, ɸ$_3^-$ are the sub states of J=1 and optically active, sitting at higher energy levels in order, ɸ$_2$ is the j$_z$=0 state with its dipole aligned in OP direction and only can be accessed with OP electric field, hence is so called "gray exciton". ɸ$_3^+$, ɸ$_3^-$ are the j$_z$=±1 states with their dipole orthogonally aligned in the quantum well plane. The four exciton states are expressed as follows:

$$\varphi^1 = \tfrac{1}{\sqrt{2}}(P_{+e}S_h\beta_e\alpha_h - P_{-e}S_h\alpha_e\beta_h)\cos\theta - \tfrac{1}{\sqrt{2}}P_{0e}S_h(\alpha_e\alpha_h - \beta_e\beta_h)\sin\theta,$$

$$\varphi^2 = \frac{1}{\sqrt{2}}(P_{+e}S_h\beta_e\alpha_h + P_{-e}S_h\alpha_e\beta_h)\cos\theta - \frac{1}{\sqrt{2}}P_{0e}S_h(\alpha_e\alpha_h + \beta_e\beta_h)\sin\theta,$$

$$\varphi^3_+ = P_{+e}S_h\beta_e\beta_h\cos\theta - P_{0e}S_h\alpha_e\beta_h\sin\theta,$$

$$\varphi^3_- = P_{-e}S_h\alpha_e\alpha_h\cos\theta - P_{0e}S_h\beta_e\alpha_h\sin\theta, \tag{1}$$

where $S = R_{60}Y_{00}$; $P_0 = R_{61}Y_{10}$; $P_{\pm} = R_{61}Y_{1\pm1}$; e and h indicate the electron and hole states, respectively; θ is a constant determined by spin orbit coupling and crystal field splitting. For an OP magnetic field applied in Faraday configuration, i.e., E⊥c; B∥c; k∥c, the magnetic field can only mix the wave function of ɸ1, ɸ2, which their dipole is along c axis and its emission can be only accessed with large numerical aperture lens under normal collection condition[12]. The eigenwave functions and the corresponding eigenenergy will be the following:

$$\emptyset^{1,2}_{Farad} = a_{1,2}\varphi^1 + b_{1,2}\varphi^2,$$

$$\emptyset^{3,4}_{Farad} = \varphi^3_{\pm},$$

$$E^{1,2}_{Farad} = \frac{1}{2}\{E_1 + E_2 \pm [(E_1 - E_2)^2 + 4\gamma^2B^2]^{1/2}\}a_{1,2}\varphi^1 + b_{1,2}\varphi^2,$$

$$E^{3,4}_{Farad} = E_3 + \delta B, \tag{2}$$

While IP magnetic field applied in Voigt configuration, i.e., E⊥c; B⊥c; k∥c, the magnetic field mix the wave function of ɸ1(ɸ2) and ɸ3⁺ or mix that of ɸ1(ɸ2) and ɸ3⁻ depending on the exciting light polarization direction is along X or Y in the quantum well plane, which is supposed to brighten DX[13,14].

$$\emptyset^{1,2}_{Voigt} = c_{1,2}\varphi^1 + d_{1,2}(\varphi^3_+ - \varphi^3_-),$$

$$\emptyset^{3,4}_{Voigt} = c_{3,4}\varphi^2 + d_{3,4}(\varphi^3_+ - \varphi^3_-),$$

$$E^{1,2}_{Voigt} = \frac{1}{2}\{(E_1 + E_3) \pm [(E_1 - E_3)^2 + 8\zeta^2B^2]^{1/2}\},$$

$$E^{3,4}_{Voigt} = \frac{1}{2}\{(E_2 + E_3) \pm [(E_2 - E_3)^2 + 8\eta^2B^2]^{1/2}\}, \tag{3}$$

where $c_{1,2} = \frac{2\zeta B}{\left[2(E^{1,2}_{Voigt}-E_1)^2+4\zeta^2B^2\right]^{1/2}}$, $d_{1,2} = \frac{E_1-E^{1,2}_{Voigt}}{\left[2(E^{1,2}_{Voigt}-E_1)^2+4\zeta^2B^2\right]^{1/2}}$,

$$c_{3,4} = \frac{2\eta B}{\left[2(E^{3,4}_{Voigt}-E_2)^2+4\eta^2 B^2\right]^{1/2}} \quad , \quad d_{3,4} = \frac{E^{3,4}_{Voigt}-E_2}{\left[2(E^{3,4}_{Voigt}-E_2)^2+4\eta^2 B^2\right]^{1/2}} \quad .$$

Many groups indicated the existence of a low energy dark state in the 2D-HOIP through indirect evidences such as the abnormal integrated PL intensity change while cooling from room temperature to 4K or lifetime measurement. Only very recently the detection of DX in static photoluminescence (PL) or magneto-absorption in 2D HOIP film is reported[15,16]. Nevertheless, those research works are far from conclusive and leave debated questions. Firstly, whether the DX emission can be observed in a pristine 2D-HOIP in the conventional PL measurement under OP magnetic field is still an open question. Do et al reported the observation of only BX at 2.34 eV in exfoliated flakes of pristine $(PEA)_2PbI_4$ single crystals, even at high OP magnetic field up to 33T. In stark contrast, Mateusz Dyksik et al.[15] and Neumann et al.[16] reported the weak emission of the DX state at ~ 2.33 eV in $(PEA)_2PbI_4$ film and the latter group even reported the DX brightening by OP magnetic field. Secondly, the energy position assigned for DX is quite different. For instance, Fang et al. attributed the main PL peak at 2.34 eV as a DX state[17] which is different from the claim of other reports. Although both Neumann et al.[16] and Dyksik et al.[15] reported the DX seated in 2.33 eV, only the latter group reports the observation of DX2($\phi_2$) and seated in 2.355 eV, even higher energy than the other two bright $\phi_3^+/\phi_3^-$ states, which is in confliction with the previous theoretical prediction[6]. Lastly, whether the DX behavior has thickness dependence or not for 2D-HOIP is still unknown, to the best of our knowledge, given that the properties of layered material are usually sensitive to the thickness.

Hereby, in this work, we performed systematic investigation of the DX evolution with different thickness of high quality exfoliated $(PEA)_2PbI_4$ single crystalline nanoflakes. We found that, for thick samples, the DX could not be detected by static PL spectroscopy under conventional configuration (E⊥c; k∥c) at our lowest temperature 4 K, as shown in Fig.1b. With samples thinning down, the DX could be gradually detected and has almost equal brightness emission compared to BX for bilayer sample with the DX emission located at 2.33 eV and the BX emission at around 2.34 eV. The separation between BX and DX is ~ 10 meV. The second important discovery of our work is that the DX of all thickness samples can be only brightened by external applied IP magnetic field but not brightened for the OP magnetic field, which is in consistence with the result of very recent work[6]. Such DX brightening can be well reconciled by the nature of the DX components, $\phi_1$, $\phi_2$ as mentioned above. Our work not only sheds light on the complex physics of such rejuvenating perovskite material but also paves a way for the exploitation of DX for the applications in opto-spintronic and quantum computing devices.

**Result and discussion:**

The $(PEA)_2PbI_4$ single crystals were fabricated by Anti-solvent Vapor assisted Capping Crystallization method[18] with normal size of half centimeter in diameter and shinning flat surfaces. The high quality of the single crystalline were confirmed by X-ray diffraction and PL spectroscopy. By exfoliation in Argon filled glovebox, tens of microns size flakes with different thickness were identified and encapsulated in hexagonal boron nitride flakes (more experimental details can be found in the Methods part). Sample thicker than 30 nm is named "bulk" sample and "few-layer" refers to samples of less than 10 layers. PL spectroscopy is performed under conventional configuration (E⊥c; k∥c). We observe very narrow PL peak in our few-layer sample with FWHM of ~ 36 meV at 260 K and ~ 5 meV at 4 K as shown in Fig.1a. On the contrary, low quality crystalline 2D perovskite manifests its defects related broad emission in PL spectrum at low energy region, as shown in Fig.S1b. In Fig.1b，for the bulk sample, only two states located at around 2.34 eV($\phi_3^-$) and 2.30 eV can be observed in the PL spectrum, which is attributed to the BX emission and biexciton (XX) emission and is consistent with most of the previous observations[8,17,19]. For few-layer sample, a new state at around 2.33eV ($\phi_1$) shows up. When thinning down to bilayer, the state at 2.33eV grows up and a new state at 2.355eV ($\phi_3^+$) can also be observed. While the peak at 2.34 eV is attributed as BX emission in most of previous work, the peak at 2.355 eV is seldomly observed and in Dyksik's work[15] it is attributed as the mixture of BX and DX of ϕ$_2$. However, for temperature dependence PL shown in Fig. 2a and magnetic field dependence shown in Fig.3b&d, it is not brightened by magnetic field or weakened with increased temperature. Therefore, we attribute it as one part of the BX. Furthermore, previous energy and time resolved PL spectra[17] demonstrated that the emission at 2.355 eV has very short lifetime, which confirms its BX nature. The temperature dependent PL spectra from 4K to room temperature show a continuous blueshift of the BX peak with increasing temperature as displayed in Fig.1d, indicating no structural phase transition, which is in accordance with previous work.

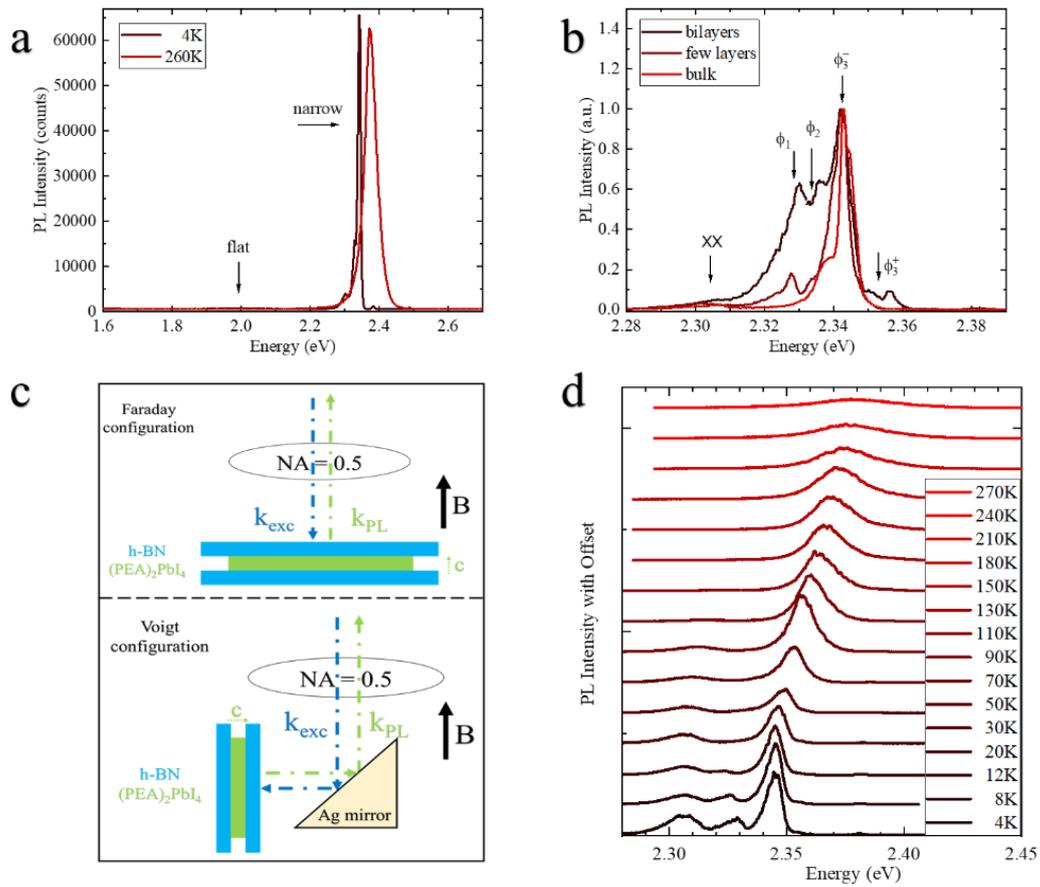

**Figure 1 Overall PL spectrum of (PEA)$_2$PbI$_4$ evolution with temperature and thickness** (a) PL spectrum at 260K and 4K of few layer sample. (b) PL spectrum for bilayer, few layer and bulk sample at 4K with higher wavelength resolution. (c) Experimental set-up for our magneto-optical measurement. Up panel: Faraday configuration; lower panel: Voigt configuration. (d) Temperature dependent PL of few layer (PEA)$_2$PbI$_4$, blue shift as temperature increases.

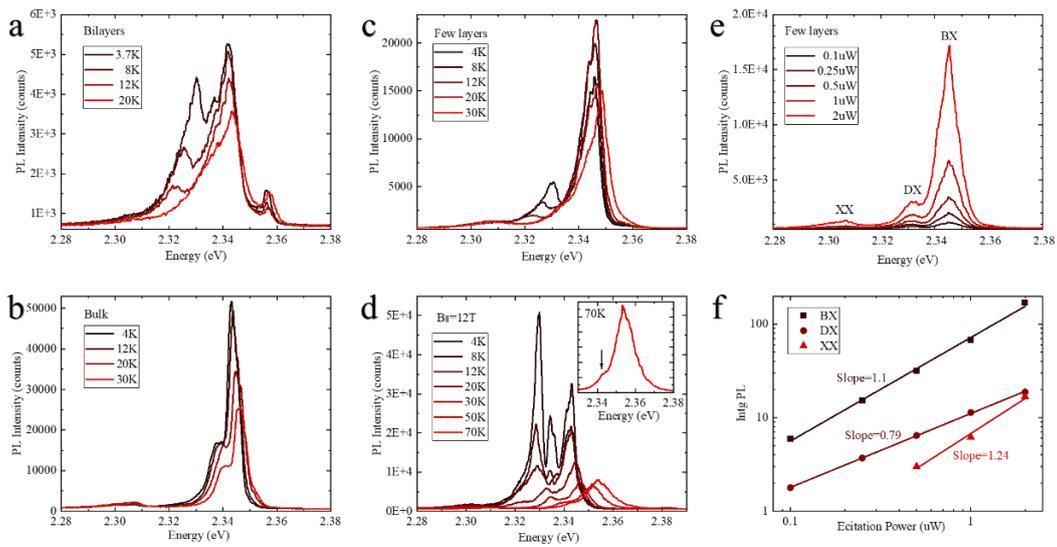

**Figure 2 Thickness dependent PL evolution with temperature and excitation power** (a)(b)(c) PL of different thickness samples. (a)Bilayers, (b)Bulk, (c)Few layers. (d) The temperature dependent PL spectrum for few-layer sample, when applied 12T IP magnetic field, the DX signature is distinguishable even at 70K. (e)The excitation power dependent PL spectrum of few-layer sample. (e) The integral PL Intensity vs excitation power. The slope for BX, DX, and XX is around 1.1, 0.79, and 1.24 respectively.

Fig.2 displays the temperature dependent PL spectrum of different thickness samples. Our results show that the ratio of DX to BX intensity grows when the thickness decreases. For the bilayer sample, the intensity of DX can be equally comparable to that of BX at base temperature, which is not reported previously. As one can see in Fig.2a, c, the intensity of DX decreases quickly with elevated temperatures, which is caused by thermal activation of the transition from DX to BX and is one of the fingerprint behaviors of DX. The laser excitation power dependent PL spectrum are plotted in Fig.2e and the power dependence for each peak is plotted in Fig.2f. One can see that the fitted power factor for peak XX is 1.24, indicating the biexciton power dependence[17]. The power factor is around 1.1 for peaks of BX, manifesting their bright nature. While for peak DX, the power factor is less than 1, which is another signature of the DX emission. Due to the more fragile and the minute size of the monolayer HOIP nanoflake, the PL spectrum of monolayer is not thoroughly studied under our current experimental condition. The acquired PL spectrum for monolayer sample is displayed in Fig. S2 and is further discussed in Supporting Information. The PL intensity of the $\phi_3^+$ state of monolayer is even higher than that of $\phi_3^-$ and does not disappear with increasing temperature, which is another evidence for its BX nature.

Our thickness dependent PL measurement at low temperatures for pristine high quality $(PEA)_2PbI_4$ samples provides unambiguous result to the above-mentioned debated question. One can confirm that the DX of pristine thick samples cannot be activated under Faraday configuration if without collecting lens with large numerical aperture[12]. For the thin samples, the activation of DX probably comes from the strain[12,15,21,22]. Strain is usually induced by the residuals existing inside the Van der Waals gaps between the perovskite sample and encapsulating BN flakes. Such unavoidable strain or ripples caused by the residual can induce the emission of DX. While the 2D HOIP resembles the monolayer TMD sample in terms of the spin related band structure and hence DX behavior, for instance, the emission of DX with dipole oriented in OP direction, can change its emission direction under strain and be collected in conventional configuration.

Another conflicting result of previous research is whether the OP magnetic field in Faraday configuration can brighten the DX or not. To clarify this question, we performed both OP magnetic field in Faraday configuration and IP magnetic field in Voigt configuration for comparison. In Fig.3, the evolution of PL spectrum versus magnetic field of both configurations of few layer and thick flakes are shown respectively. One can notice that the DX can be brightened in the Voigt configuration while the DX emission of few layer and bulk samples are hardly changed by the OP magnetic field in the Faraday configuration. From the peak intensity statistic plot versus IP magnetic field as shown in Fig.3f, one can easily notice that, IP magnetic field enhances the DX PL peak gradually with increasing of the field strength while the BX intensity gradually decreases. This observation of DX behavior under different magnetic field configuration can be well explained with the picture of different magnetic field configuration mix different components of exciton wave function[12,15,21,22], which is also confirmed very recently in a IP magnetic field absorption and PL experiment[15].

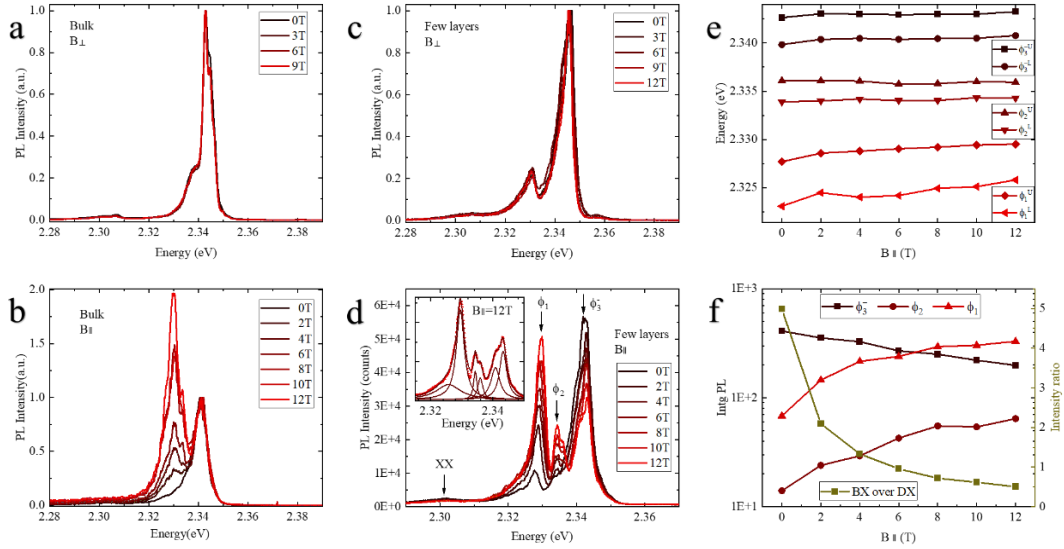

**Figure 3. OP and IP magnetic field dependent PL.** (a) (c)Bulk and few layers under OP magnetic field. (b)(d) Bulk and few layers under IP magnetic field. Inset of (d): Lorentz fit of the $\phi_1$, $\phi_2$, $\phi_3^-$ states under 12T IP magnetic field. (e) The fitted subpeak energy versus IP magnetic field. (f) The PL Intensity of each state and the Intensity ratio of BX($\phi_3^-$) to DX($\phi_1+\phi_2$) extracted from (d).

The inset in Fig.3d, shows that $\phi_1$, $\phi_2$, $\phi_3^-$ has two apparently splitting subpeaks, which can be well fitted with two Lorentz shape curves for few layer samples. The fine structure splitting for $\phi_3^+$ can be seen in bilayer and monolayer samples, displayed in Fig.1c and Fig.S2,S3. The fine structure splitting is only reported for $\phi_3^-$ state previously in exfoliated high quality single crystalline sample[19]. As shown in Fig.3e, our results demonstrate that all the four states have a large 0 Tesla splitting of 2 – 4 meV due to the large e-h exchange interaction resulted from the 2D quantum well structure. Such large e-h exchange splitting surpasses the spin energy splitting of our maximum magnetic field 12 Tesla, therefore the magnetic field dependent energy curves are not able to show any noticeable Zeeman splitting.

We also performed time-resolved PL (tr-PL) measurement on $(PEA)_2PbI_4$ few layer samples. Fig.4a, shows the time resolved PL spectra evolution with IP magnetic field, which has two decay components, a fast one has lifetime around 100 ps, and a slow lifetime around tens of nanoseconds. We attribute the former short decay time to the BX lifetime and the latter longer one to the DX lifetime. The extracted data is shown in **F**ig.4b. The ratio of the slow part to the fast part increases with IP magnetic field, which is consistent with the static PL evolution, and verifies that IP magnetic field can mix the BX and DX state. Fig.4d shows the tr-PL spectra as the temperature increases under 12 Tesla IP magnetic field. The slow component does not totally disappear until 70K compared with Fig.4c, which has zero magnetic field, but indeed declines. This is also consistent with the static PL results when temperature increases.

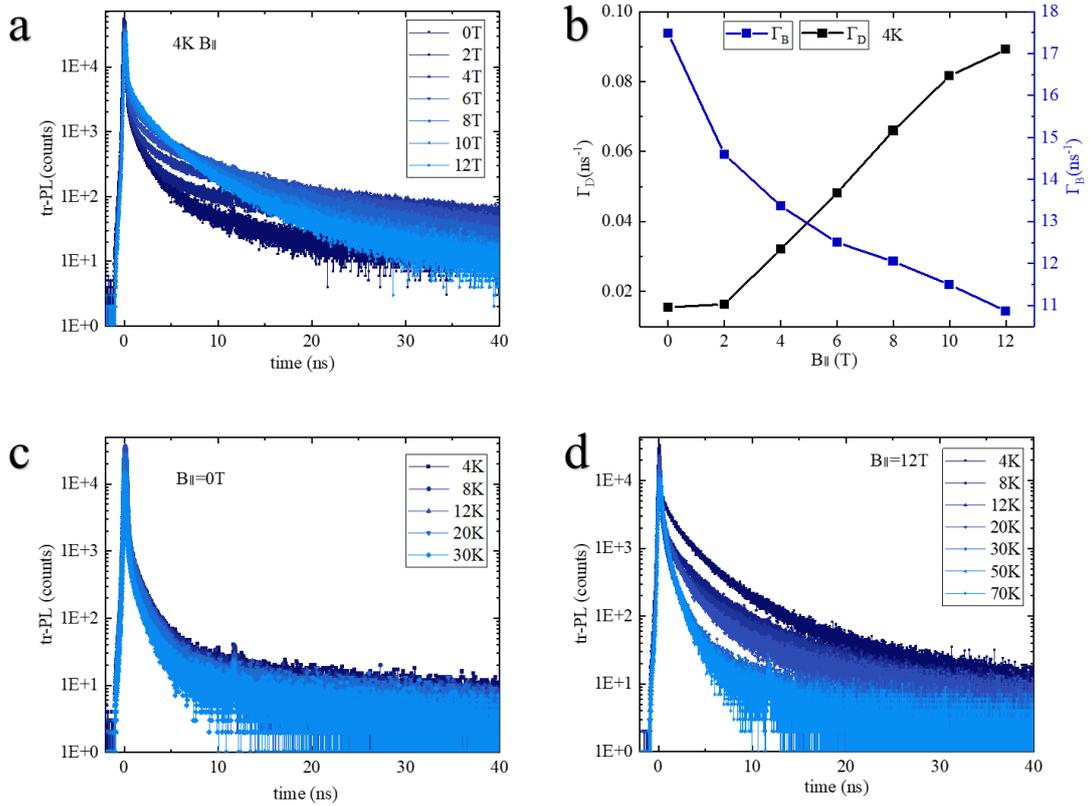

**Figure 4. Time resolved PL spectroscopy** (a) under different IP magnetic field. (b) The low decay rate ($\Gamma_D$) and fast decay rate ($\Gamma_B$) extracted from (a). (c) 0-30K Temperature dependence without magnetic field. (d) 0-70K Temperature dependence under 12 Tesla IP magnetic field.

In conclusion, we investigate the DX emission behavior by magneto PL

spectroscopy on mechanically exfoliated (PEA)$_2$PbI$_4$ single crystal samples with different thicknesses. The DX emission is detected for pristine bilayer and few layer samples at low temperature but not for thicker samples which we attribute to the strain induced DX activation in thin samples. The key finding is that OP magnetic field applied in Faraday configuration cannot brighten the DX in all thickness samples. With the IP magnetic field brightened emission, sub component of DX ϕ$_2$ at 2.334 eV and the fine splitting of all the four exciton states are observed in static PL spectroscopy for the first time. Our report of such thickness and magnetic field direction dependent DX behavior clarifies the debating questions in the halide perovskite DX research community. Furthermore, the discovery of brighten of DX in thin samples without the aid of magnetic field and its tunability shades light on the application of halide perovskite material on the opto-spintronic and quantum computing devices.

**Methods:**

**1. 2d halide perovskite single crystal preparation:**

**Materials:** -Butyrolactone (GBL, 99.9%) was purchased from Aladdin Corp. Phenylethylammonium iodide (PEAI, 99.5%) and Lead (Ⅱ) iodide (PbI$_2$, 99.99%) were purchased from Xi'an Polymer Light Technology Corp. All solvents and reagents in the experiment were used directly without purification.

**(PEA)$_2$PbI$_4$ Single Crystal Fabrication:** The (PEA)$_2$PbI$_4$ precursor solution (1.0 M) was prepared by mixing 0.248 g PEAI (1.0 mmol) and 0.2305 g PbI$_2$ (0.5 mmol) in 0.5 mL GBL. The solution was then stirred at 50 °C overnight in N$_2$ glovebox and filtered with a 0.22 μm PTFE syringe filter.

The (PEA)$_2$PbI$_4$ single crystals were fabricated by "Anti-solvent Vapor assisted Capping Crystallization (AVCC)" method as reported previously[18]. Specifically, 3×3 cm$^2$ glass substrates (thickness ~ 1 mm) were first cleaned consecutively in isopropanol, acetone, acetone, isopropanol ultrasonic bath for 10 min, respectively, and then dried in an oven at 60 °C. After that, 30 μL of the filtered precursor solution was drop onto the precleaned glass substrate placed in the middle of a glass petri dish, and quickly capped by another glass substrate. Then, four small containers with 2 mL DCM solvent were placed around the glass substrate. Finally, the cover of the petri dish was placed atop to form an airtight environment. In about six hours, the rectangle-shaped crystals can be collected between the two substrates. The above crystal growth process was all carried out in a fume hood.

## 2. AFM&XRD

The thicknesses of the sample were measured on non-contact mode by Park NX10 Atomic Force Microscopy (Park systems, Korea).

The X-ray diffraction experiment was done on a single crystal at room temperature. (Empyrean, Malvern Panalytical Ltd)

## 3. Magneto photoluminescence measurement:

Our sample was set in a closed cycle cryostat (Teslatron PT, Oxford Instruments Inc.) with a maximum magnetic field 12T. The magnetic field direction and the position of the optical window was fixed, so we use an Ag mirror inside the cryostat to change the direction of light and so the sample relative position to achieve the Voigt configuration. Using a 50x objective lens (LMPLFLN50x, Olympus, NA=0.5) for the excitation and collection. A 473nm continuous wave laser (Changchun New Industries Optoelectronics Technology Co. Ltd) for the excitation of PL spectrum, and a 450nm picosecond laser diode (pulse width 20ps, Pico Quant,) for both PL and tr-PL measurements. The optical signal was dispersed by a single grating 1800-(150-) grooves/mm (300mm) spectrometer (SpectraPro HRS300, Princeton Instruments Inc.), and recorded by a nitrogen-cooled CCD (PyLoN, Princeton Instruments Inc.) for the PL spectrum. Time-resolved photoluminescence measured using the same experiment set-up, the signal was detected by an avalanche photodiode (IDQ), and counted by a TCSPC Module (PicoHarp 300, Pico Quant), with a time resolution of 4ps.

## Acknowledgements

L.J Li acknowledges the funding from National Key R&D Program of China (2019YFA0308602), the state key program (91950205) and the general program (11774308) of National Science Foundation of China, the Zhejiang Provincial Natural Science Foundation of China (LR20A040002), Y.J Fang acknowledges the funding of National Natural Science Foundation of China (No. 52003235 and No. 62075191). D.R Yang acknowledges the funding of National Natural Science Foundation of China (No.62090030 and No. 61721005).

## Contributions

W.T. performed characterization of the single crystal, preparation of the nanoflakes and the magneto-optical measurements.  L.T. and Y.F. grew the single crystals, T.Z. helped on constructing the optical setup. W.T. and L.L analysed data and wrote the manuscript with input from all other authors.

Wei Tang: http://orcid.org/0000-0002-4641-1855

Liting Tao: http://orcid.org/0000-0001-7278-1115

Tian Zhang: http://orcid.org/0000-0002-8475-7208

Yanjun Fang: http://orcid.org/ 0000-0002-1745-2105

Deren Yang: http://orcid.org/ 0000-0002-1745-2105

Linjun Li: http://orcid.org/0000-0002-2734-0414

## Ethics declarations

Competing interests

The authors declare no competing interests.

# Supporting Information

# Thickness dependent dark exciton emission in (PEA)$_2$PbI$_4$ nanoflake and its brightening by in-plane magnetic field

Wei Tang [1], Liting Tao[2], Tian Zhang[1], Yanjun Fang [2*], Deren Yang[2], Linjun Li [1*]

1 State Key Laboratory of Modern Optical Instrumentation, College of Optical Science and Engineering,

2 State Key Laboratory of Silicon Materials, School of Materials Science and Engineering,

Zhejiang University, Hangzhou 310027, China.

*Corresponding authors: lilinjun@zju.edu.cn; jkfang@zju.edu.cn

1. **Sample characterization and quality discussion**

The (PEA)$_2$PbI$_4$ single crystal is of high quality which can be confirmed by the X-ray diffraction result as shown in Fig.S1a and the microscopic picture shown in Fig.S1c, nanoflakes derived by mechanical exfoliation are tens of micrometers in lateral size with different thicknesses ranging from ~100 nm to few layers and even monolayers. AFM shows flat and homogeneous surfaces for the thick flakes and a few scattering points distributed in the thin flakes which have similar color with the thick flake (Fig.S1d), which maybe the PDMS residual or trapped air bubbles. Those differences between thick and thin samples are probably the source for the strain we discussed in the main text. And one can see the nanoflake without a top h-BN degenerate severely once exposed to air even just a few minutes (Fig.S1c, bottom right). But flake fully encapsulate can keep its high quality even contacts with water drop. Fig.S1b displays the difference of PL spectrum between high quality and low quality samples as mentioned in main text, which the higher quality sample does not show an obvious broad hump ranging from 1.6 eV to 2.2 eV in PL spectrum. This broadband emission usually is attributed to self-trapped exciton[1].

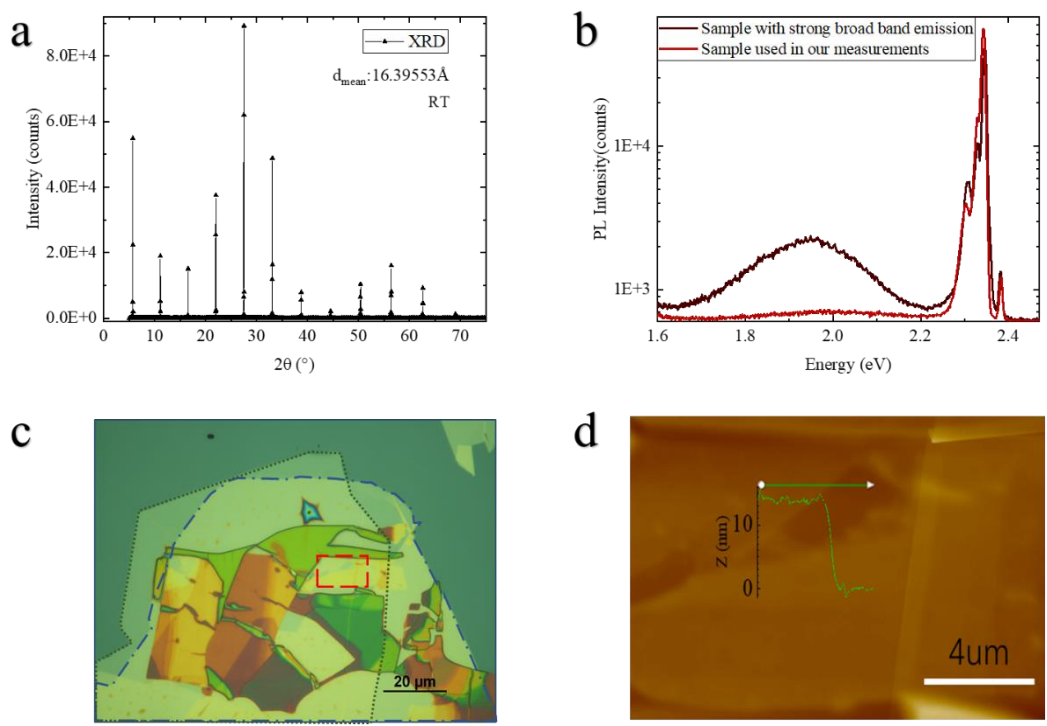

**Fig.S1 Characterization of the (PEA)₂PbI₄ sample** (a) X-ray diffraction pattern of single crystal (PEA)₂PbI₄. The distance along c axis is around 16.4 Å. (b) PL spectrum with(dark red) and without(red) strong broad band emission (attribute to self-trapped exciton). (c) Optical image of one representative sample encapsulated by bottom- (blue dotted dash line), and top- (green dotted line) h-BN. (d) AFM image of red dash line district in (c)，the green line on (d) indicate a typical few-layer sample around 14nm.

2. The PL result of monolayer sample and discussion

From Fig.S2 one can see the monolayer (PEA)₂PbI₄ nanoflake show similar PL pattern at low temperature to that of bilayer and few layers sample, but very different from that of bulk sample which are already shown in main text. The most obvious feature is that the peak around 2.35 eV ($φ_3^+$) has higher intensity than the lower energy part of the BX at 2.34 eV ($φ_3^-$). Both the two features survive till high temperature, which shows their bright nature. As shown in Fig. S2c, the exciting power dependence of emission of $φ_3^+$ and $φ_3^-$ states gives the linear dependence coefficient of around 1, while the dark state $φ_1$ dependence coefficient is 0.78, less than 1. This confirms further the emission at 2.35 eV is of BX nature, as pointed out in main text. In Fig.S2d, the OP magnetic field dependence of PL of monolayer is plotted. One can see, the intensity of all the states are slightly enhanced by OP magnetic field, including the DX, which is not conflicting with the conclusion that OP magnetic field cannot brighten DX since in thin samples, the strain induced c axis rotation could make the magnetic field has an IP component and therefore can also enhance the DX. Overall, more investigations are needed for unveiling the underlie intrinsic physics that governing the monolayer behavior.

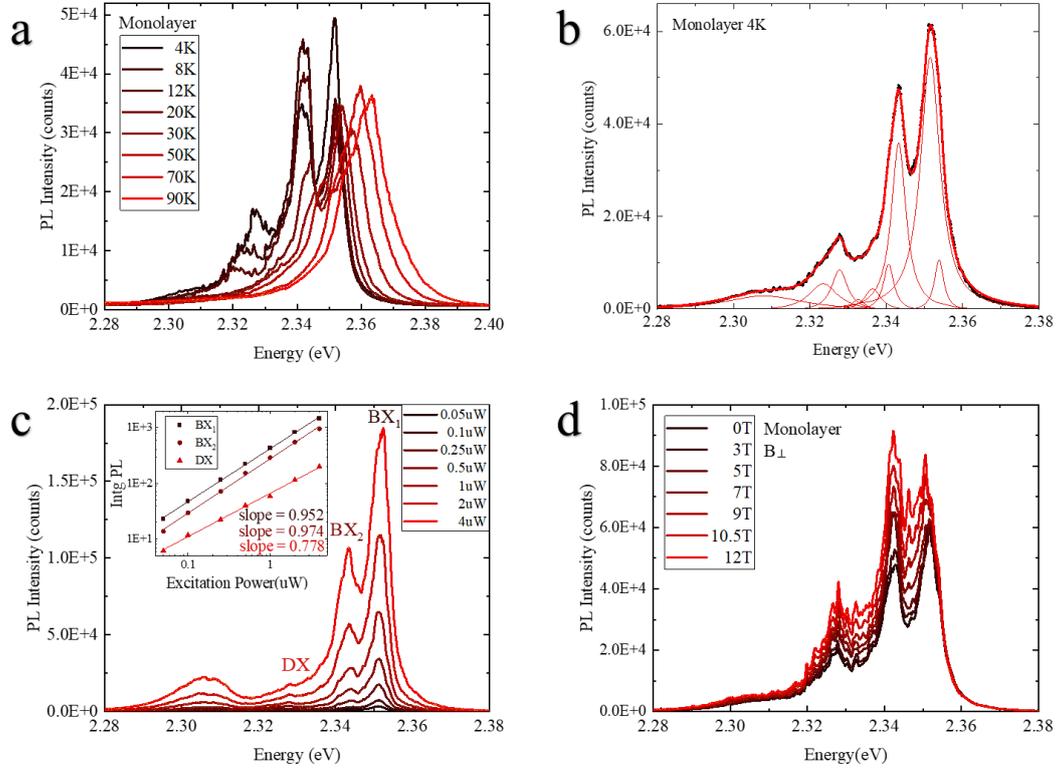

**Fig.S2 PL spectrum of monolayer (PEA)$_2$PbI$_4$ nanoflake** (a) Temperature dependent PL from 4K to 90K. (b) Lorentz Fit of the 4K spectrum, ɸ$_3^+$ 、ɸ$_3^-$ 、ɸ$_2$ 、ɸ$_1$ are all fitted with two Lorentz shape line. (c) Power dependent PL. Inset: The linear fit of integrated PL intensity vs excitation power, the slope of BX$_1$, BX$_2$ and DX state is 0.952, 0.974, and 0.778 respectively. (d) PL under OP magnetic field.

## 3. The PL under OP magnetic field of bilayer sample and discussion

Fig.S3a displays the PL result of bilayer sample under OP magnetic field. One can see that similar to the monolayer sample, OP magnetic field can also brighten DX in bilayer sample. This confirms that in thin flakes, the strain probably rotates the electric dipole direction of the exciton states, so that the OP magnetic field would have a projecting IP magnetic field component. Fig.S3b shows the circular polarization dependent PL under increasing OP magnetic fields. With increasing $B_\perp$, the degree of circular polarization (DCP) increases, confirming that OP magnetic field tilts the IP spin to OP direction gradually. DCP = $(\sigma_+ - \sigma_-)/(\sigma_+ + \sigma_-)$. DCP is calculated to be 21.4%, 9.4%, -5.8% for $\phi_3^+$, $\phi_3^-$, $\phi_1$ respectively. The difference of DCP between these three states may result from the coupling between them enabled by magnetic field while the relatively low DCP value is because that the e-h exchange interaction is strong in this 2D compound as also discussed in previous work[2,3].

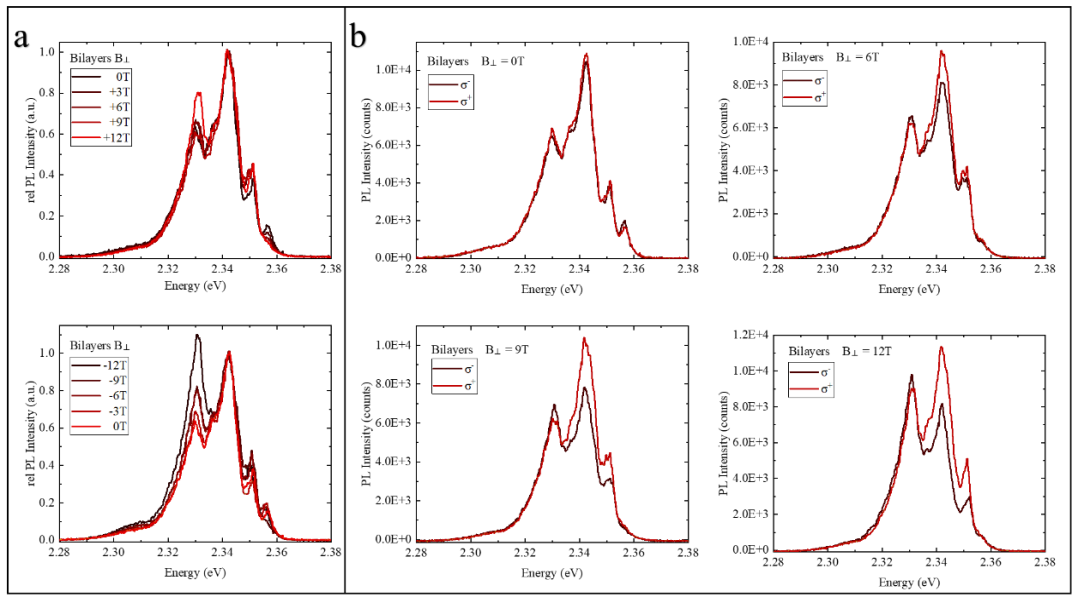

**Fig.S3 PL spectrum of bilayer $(PEA)_2PbI_4$ nanoflake** (a) PL under OP magnetic field. Positive and negative means the direction of B is opposite, but both parallel to c axis of the $(PEA)_2PbI_4$ flake. (b) Circularly polarized resolved $\sigma^-$ (green) and $\sigma^+$ (blue) PL spectra at different OP magnetic field. The circular polarized degree is increase when applied OP magnetic field.

4. **Time resolved PL spectrum fitting and discussion**

In Fig.S4a&b, the time resolved PL decay curves for thick and few-layer sample are fitted with ExpDecay2 function respectively and plotted, where the inset plot shows the corresponding static PL spectrums. One can see, each PL decay curve can be fit quite well and it gives out one fast component of ~ 100 ps and one slow component of ~ 1 ns as plotted in Fig.S4d for few-layer sample result. The temperature dependent PL lifetime demonstrates two sections. One section is the high temperature ranging from 130 K to room temperature, where the slow and fast components of PL lifetime have the same temperature dependence, while in the other section ranging from 130 K to lowest temperature 4 K, fast and slow components have opposite temperature dependence. This seemingly complex temperature dependence of lifetime, can be well explained by the fact of changing on exciton phonon coupling strength at the temperature 130 K. In Fig.S4c, the full width at half maximum (FWHM) value of the BX peak for few layer sample is derived from Fig.1d in the main text and plotted versus temperature. It can be fit with the equation of $\Gamma(T) = \Gamma_0 + \sigma_{Ac}T + \Gamma_{LO}[\exp(E_{LO}/k_BT)-1]$, where Γ(T) is the ZPL linewidth at temperature T and 0 its value at zero temperature, $k_B$ is the Boltzmann constant, $\sigma_{Ac}$ and $\Gamma_{LO}$ are, respectively, the exciton–acoustic phonon and the exciton–longitudinal optical (LO) phonon coupling coefficients, and $E_{LO}$ is an effective optical phonon energy representing the carrier–LO phonon interaction[4,5]. The fitting apparently can be divided into two temperature regions separated at 130 K. The fitting of FWHM data at high temperature region gives out strong coupling between exciton and optical phonon. In stark contrast, for the low temperature region, the exciton optical phonon coupling strength is only ~ 1/30 of that high temperature region. We can now see that how the dependence of fast and slow part of PL lifetime on temperature forms. The fast component of lifetime corresponds to lifetime of BX while the slow part corresponds to that of BX coupling with DX states. At high temperatures, because of dominating by strong coupling between exciton phonon coupling, the slow and fast part show almost same temperature dependence. At low temperatures, the coupling between exciton phonon gets weak, the coupling between BX and DX themselves will dominate the temperature dependent lifetime evolution. One can see, when such coupling becomes strong at around 70 K, where DX starts borrowing PL spectrum weight from BX, the lifetime of BX decreases and that of DX increases.

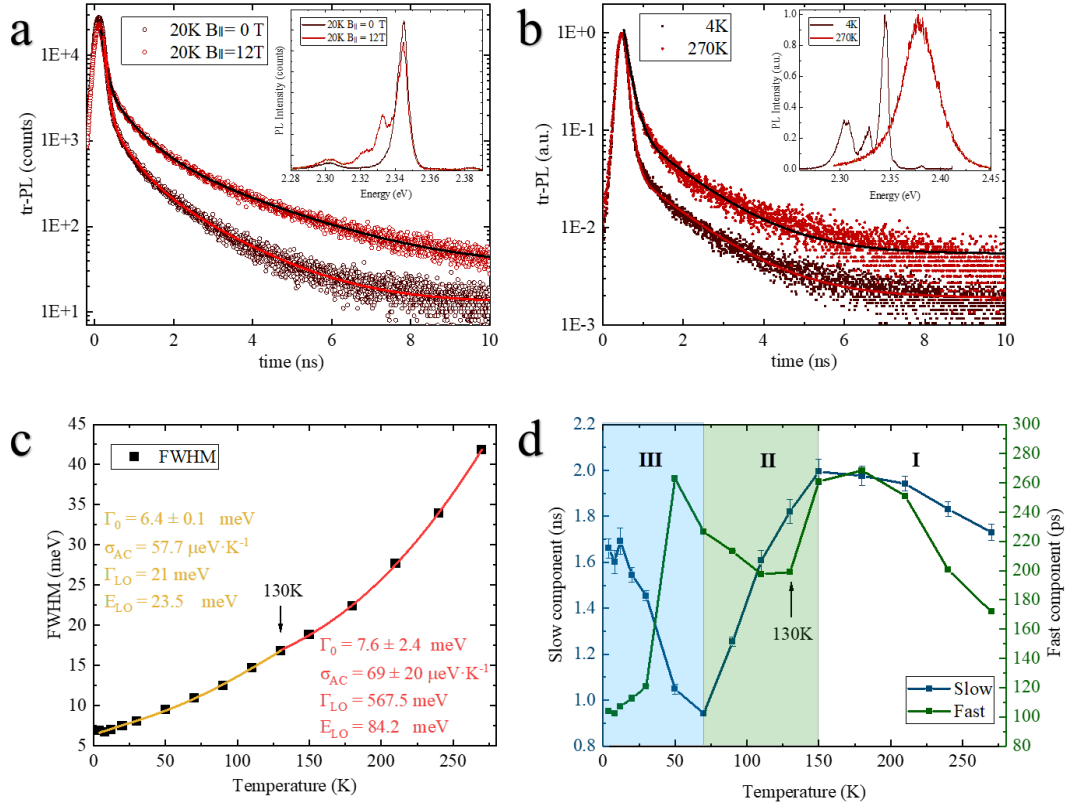

**Fig.S4 PL lifetime fitting and temperature dependence** (a) Time resolved PL at 20K under 0T (no noticeable DX emission) and 12T (with distinguishable DX emission) IP magnetic field. Inset: PL spectra under the same condition. (b) Time resolved PL at 4K and 270K respectively. Inset: PL spectra under the same condition. (c) Full width at half maximum of bright exciton PL spectra vs temperature, and fitted with equation $\Gamma(T) = \Gamma_0 + \sigma_{Ac}T + \Gamma_{LO}[\exp(E_{LO}/k_BT) - 1]$. (d) The fast and slow component of tr-PL vs temperature, fitted with ExpDecay2 equation. Ⅰ: 150K-RT; Ⅱ: 70-150K; Ⅲ: 0-70K.